\documentclass[twocolumn,aps,pre,showpacs]{revtex4}
\usepackage{dcolumn}
\usepackage{bm}
\usepackage{graphicx}

\begin{document}
\title{Effective charge, collapse and the critical point of a polyelectrolyte chain}
\author{Arindam Kundagrami and M. Muthukumar$^{*}$}
\affiliation{Department of Polymer Science and Engineering,
University of Massachusetts, Amherst, MA 01003, USA}
\begin{abstract}
The charge of a polyelectrolyte (PE) controls myriads of phenomena in biology, biotechnology, and materials
science, but still remains elusive from an understanding.
Considering the adsorption of counterions on an isolated PE chain,
an analytical expression for the effective charge
- valid for all chain flexibility,
for variable salt, in
good solvents at all conditions, in poor solvents in the expanded (coil) state away
from the critical point, and for gels, brushes and other PE systems
in their expanded state - is derived. Phase boundaries and the critical point
for the I-st order collapse transition, induced cooperatively by counterion adsorption and
chain conformations, are calculated self-consistently. 
The size of the PE chain is found to be a single-valued
function of charge.
\end{abstract}
\maketitle

The net charge of a polyelectrolyte (PE) chain in salty aqueous solutions is the most fundamental albeit 
elusive entity faced in systems containing charged polymers.  
Manifestations of charged polymers, whether m-RNA, ss- or ds-DNA, proteins, polysaccharides, or 
synthetic analogs such as polystyrene sulfonates, and the ways in which they fold conformationally and move 
in solutions depend crucially on their charge. These conformational fluctuations of PE chains
under external stimuli are of critical importance to numerous biological and synthetic processes. A charged
chain assumes in a good solvent a much larger
dimension\cite{bee-stekre-win-liumut-97-95-98-02,mut-04} than an uncharged self-avoiding-walk (SAW) 
chain due to electrostatic repulsion between like-charged monomers. This
repulsive force resists the collapse of the PE in a poor solvent, but only up to a certain degree of hydrophobicity, beyond
which it assumes similar to neutral chains a dimension way below the Gaussian value.\cite{wiletal-81,mut-84}
Additionally, the PE has its counterions some of which adsorb
(i.e., condense) on the monomers and modulate the effective charge. An optimization between
electrostatic binding energy and translational entropy controls this effective charge which, in turn, affects the size of the chain 
and vice versa. Therefore, a self-consistent procedure is necessitated to calculate the size and the effective charge
of the PE molecule.

A number of theories\cite{sheyos-hathi-duavil-sheyet-schall-07-92-05-99} addressed
the size dependency of a PE
chain (or a PE gel\cite{tanetal-80}) on its effective charge, which is taken as
a fixed parameter independent of its size.
However, a fixed finite charge is incompatible
to a collapsed PE chain in poor solvents as it leads to 
severe electrostatic energy penalty.
Variable PE charge was addressed\cite{dobrub-01} within the Oosawa model assuming rod-like chains.
A three-state model\cite{khoall} earlier self-consistently calculated the size and charge with the
free counterions partitioned into two domains - one within and the other outside the
radius of gyration ($R_g$) of the PE. A successful analysis\cite{brietal-98} of collapse due to counterion correlations 
in good solvents, however, 
failed to include the counterion adsorption effects and the right experimental trends. 
A theory that captures counterion adsorption,
intra-chain electrostatic repulsion, and polymer 
conformational changes in a self-consistent way is yet to emerge. 

Experiments\cite{yosexptall,lohetal-08} and simulations\cite{yetall} have indisputably demonstrated
that a PE chain at a specific poorness
of the solvent collapses to a globule. However, the stability of the exciting intermediate pearl-necklace
phase\cite{dobetal-lyuetal-limall-kirall-96-99}, made of globules
and elongated chain parts co-existing in a single chain, remains a controversial\cite{yetall} issue.
In this letter we
analyze the variational theory of counterion adsorption\cite{mut-04,mut-87}, and derive
an analytical expression for the charge of a flexible PE chain in good solvent conditions.
Further, we
determine the condition for collapse of the PE chain
in a poor solvent and calculate the phase boundaries for this first-order coil-globule transition
induced {\it cooperatively} by counterion adsorption and solvent poorness. Temperature  
controls both the solvent quality and the Coulomb strength.
This formalism, fundamentally different from the Manning condensation theory\cite{manning-69}, 
accounts for the interplay between chain conformations and 
counterion adsorption from a neutralizing plasma, as well as for the ``oily" chemical 
nature of the polymer backbone. Addressing such interplay is crucial to understand the  
transformations between expanded and contracted polymer conformations encountered in a multitude of life processes, 
and to further explore a possible
decoupling of charge and size. The simplest candidate of coil-globule transition to
exhibit the new paradigm is used to elucidate ideas pertinent to the swelling of ionic gels, behavior of
charged polymers at surfaces and interfaces, and assemblies such as viruses. 

The total free energy of the system consisting of an isolated flexible
PE chain with ionizable monomers, counterions,
and the added salt ions (all monovalent) in high dilution
is taken to be\cite{mut-04} $F=\sum_iF_i$ where,
\begin{eqnarray}
\frac{F_1}{N k_B T}&=&\left(f_m-\alpha\right)
\log\left(1-\frac{\alpha}{f_m}\right)
+\alpha \log\left(\frac{\alpha}{f_m}\right), \nonumber \\
\frac{F_2}{N k_B T}&=&\left(f_m-\alpha+\frac{\tilde{c}_{s}}{\tilde{\rho}}\right)
\log\left\{\tilde{\rho}\left(f_m-\alpha\right)+\tilde{c}_{s}\right\} \nonumber \\
&+&\frac{\tilde{c}_{s}}{\tilde{\rho}}\log\tilde{c}_{s}
-\left(f_m-\alpha+2\frac{\tilde{c}_{s}}{\tilde{\rho}}\right), \nonumber \\
\frac{F_3}{N k_B T}&=&-\frac{2}{3}\sqrt{\pi}\tilde{l}_B^{3/2}
\frac{1}{\tilde{\rho}}\left\{\tilde{\rho}\left(f_m-\alpha\right)
+2\tilde{c}_{s}\right\}^{3/2}, \nonumber \\
\frac{F_4}{N k_B T}&=&-\alpha \tilde{l}_B \delta, \nonumber \\
\frac{F_5}{N k_B T}&=& \frac{3}{2 N}\left( \tilde{l}_1 - 1 - \log \tilde{l}_1 \right)
+ \frac{4}{3}\left( \frac{3}{2 \pi} \right)^{3/2} \frac{w}{\sqrt{N}} \frac{1}{\tilde{l}_1^{3/2}} \nonumber \\
&+& \frac{w_3}{N \tilde{l}_1^3}
+ 2 \sqrt{\frac{6}{\pi}} f^2 \tilde{l}_B \frac{N^{1/2}}{\tilde{l_1}^{1/2}} \Theta_0 (a), \qquad \mbox{where} \nonumber \\
\Theta_0 (a)&=&\frac{\sqrt{\pi}}{2}\left( \frac{2}{a^{5/2}}
- \frac{1}{a^{3/2}} \right) \exp (a) \mbox{erfc} (\sqrt{a}) + \frac{1}{3 a}
+ \frac{2}{a^2} \nonumber \\
&-& \frac{\sqrt{\pi}}{a^{5/2}} - \frac{\sqrt{\pi}}{2 a^{3/2}},
\label{Freen}
\end{eqnarray}
with the contributions being due to, respectively, the entropy of mobility along the chain backbone of adsorbed 
counterions ($F_1$), the translational entropy of the free ions ($F_2$), fluctuations in densities of all mobile ions
($F_3$, as in the Debye-H\"{u}ckel theory), adsorption (Coulomb) energy of the bound pairs ($F_4$), and the 
free energy of the chain\cite{mut-87} ($F_5$) resulting from its connectivity (conformational entropy), the excluded volume
(two-body), steric (three-body), and screened electrostatic interactions between monomers.
Here $f_m\equiv N_c/N$, $\alpha\equiv M/N$, with $N, N_c$ and $M$ being the numbers, respectively,
of monomers, of ionizable monomers, and of adsorbed counterions, $k_B$ the Boltzmann constant, 
and $T$ the temperature. $\tilde{c}_{s}\equiv c_sl_0^3 \equiv n_sl_0^3/\Omega$, and
$\tilde{\rho}\equiv \rho l_0^3 \equiv Nl_0^3/\Omega$ with $n_s, l_0$ and $\Omega$ being,
respectively, the number of salt cations, the length of a monomer (Kuhn step length for a
flexible polymer), and the volume of the
system. $\tilde{\kappa}=\sqrt{4\pi\tilde{l}_B\left\{\tilde{\rho}\left(f_m-\alpha\right)
+2\tilde{c}_{s}\right\}}$ is the dimensionless inverse Debye length.
Here, $\tilde{\kappa}=\kappa l_0$, and $\tilde{l}_B=C_1/T$ is the dimensionless Bjerrum
length with $C_1=e^2/4\pi k_B \epsilon_0 \epsilon l_0$, where $e, \epsilon_0$, and
$\epsilon$ are the electron charge, vacuum permittivity, and the solvent dielectric
constant, respectively. $\delta = (\epsilon l_0/\epsilon_l d)$ with $\epsilon_l$, and $d$
being, respectively, the local dielectric constant\cite{dieall,mut-04}, and the ion-pair separation.
Further, $\langle R^2 \rangle = N l l_1 \equiv N l^2 \tilde{l}_1 = 6 R_g^2$ with,
$\tilde{l}_1=l_1/l$ and $f=(N_c-M)/N=f_m-\alpha$.
$\Theta_0 (a)=2/15 (a\ll 1), 1/3a (a\gg 1)$ is a cross-over function\cite{mut-87,mut-04} 
of $a \equiv \tilde{\kappa}^2 N \tilde{l_1} / 6$. 

Generally, $F=\sum_iF_i$ must be minimized
self-consistently with respect to both $f$ and $\tilde{l}_1$ to obtain the equilibrium
values of the variables. We note that $f$ features in $F_5$ only in the
fourth term that describes the long-range electrostatic interactions between charged monomers.
However, the adsorption energy ($F_4$) related to the short-range ion-pair electrostatic attraction
is more significant to $F_5$ and shows a greater variation with $f$ in most conditions. 
At low salt both the logarithmic and the linear
terms in $F_2$ will dominate the $f^{3/2}$ term in $F_3$, and at high salt the
variation of $F_3$ with $f$ is negligible. The full numerical analysis confirms these predictions,
and allows us to 
propose an {\it adiabatic approximation} in which  the chain free energy ($F_5$)
is decoupled from the rest. Then $F_{\mathrm{ad}}=F_1+F_2+F_4$ are
the relevant contributions that determine the effective charge $f$. 
This adiabatic approximation, however,  progressively fails for lower molecular weights, higher
$l_B$'s and lower $\delta$'s ($F_5$ becomes important and comparable to $F_4$ in those cases), but, otherwise,
remains applicable for a reasonably wide range of experimental conditions, especially at low and very high salts.
Within adiabatic approximation, therefore, $F_{\mathrm{ad}}$ does not contain the expansion factor and we can minimize over
only $\alpha$ to obtain the effective charge. $\partial F_{\mathrm{ad}}/
\partial \alpha = 0$ gives us, using $f=f_m-\alpha$,
\begin{eqnarray}
f=\frac{-\left(\tilde{c}_{s}+e^{-\delta \tilde{l}_B}\right)
+\sqrt{\left(\tilde{c}_{s}+e^{-\delta \tilde{l}_B}\right)^2
+4\tilde{\rho}f_m e^{-\delta \tilde{l}_B}}}{2\tilde{\rho}}
\label{charge}
\end{eqnarray}

\begin{figure}
\includegraphics[width=8.2cm]{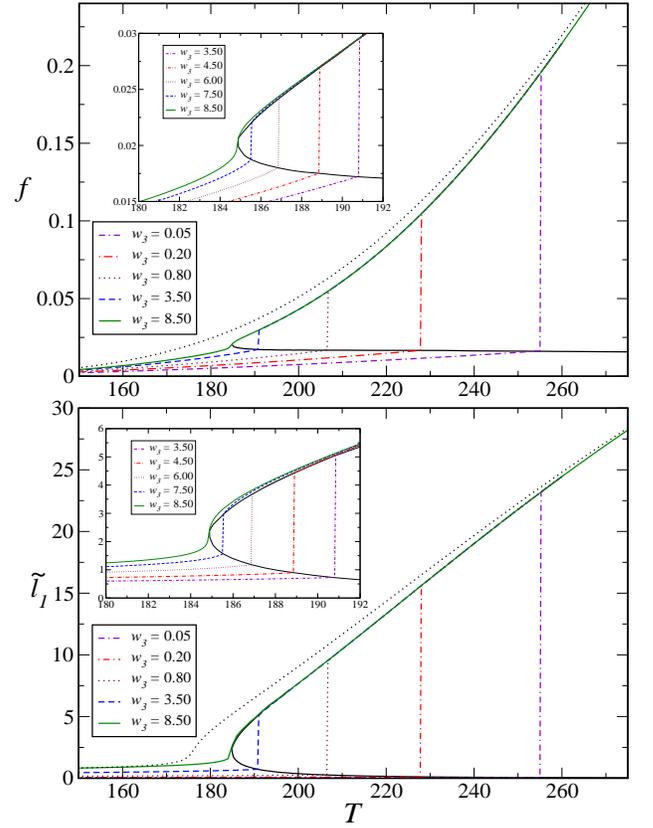}
\caption{\label{fig1}Temperature induced collapse, phase boundaries for the chain charge and size, 
and the critical point of a fully ionizable polyelectrolyte
chain. Fixed parameters are:
$N=1000$, $\tilde{\rho}=0.0005$, $\delta=3.0$, Ionizability $f_m=1.0$, $\Theta_t =400$K,
and $\tilde{c}_{s1}=0.0$. Critical values are: $T^\star=184.9$K, $w_3^\star=8.5$, $\tilde{l}_1^\star=2.3$,
$f^\star=0.020$, and from $T^\star$, $\tilde{l}_B^\star=4.867$, $w^\star=-1.163$. Dotted curves are the
analytical result [Eq. (\ref{charge})] in the expanded state.}
\end{figure}

\noindent ($f_m=1$ for a fully ionizable chain). The present model is immediately related to experimental systems, 
where the ``oily" backbone of the polymer presents heterogeneities in the dielectric function of the medium.
Eq. (\ref{charge}) is the expression for the effective charge of an isolated
PE chain, regardless of its flexibility, at all conditions in a
good solvent, and in the expanded (coil) state
away from the critical point in a poor solvent. This closed form
analytical expression, as a function of the monomer density, maximum ionizability,
salt concentration, the Bjerrum length (temperature and the bulk dielectric constant),
and the dielectric mismatch parameter $\delta$, is reported for the first
time in literature. $\delta=C_2\epsilon l_0$ has a non-universal constant
$C_2=(\epsilon_l d)^{-1}$, which can be determined from experiments using
Eq. (\ref{charge}). Further, the contributions to $F_{\mathrm{ad}}$
only depend on the number of ionizable monomers inside a polymeric system,
not on the size and shapes of the chains. In addition, the effective charge depends only on
the average monomer density $\rho$ (in volume $\Omega$), but not explicitly on the
molecular weight ($N$).
The generality of the effective charge expression warrants
its validity for
gels, brushes, and any other PE system in
dilute solutions, as long as with changing $f$ the variation in the adsorption energy for bound
counterions ($F_4$) is more significant to the variation of the intra-chain 
electrostatic interaction energy (term 4 in $F_5$). For denser systems,
higher counterion adsorption is expected in order to minimize the intra-chain
repulsive energy. 
The size of the chain in the adiabatic (or expanded) state
is determined by minimizing $F_5$ with respect to $\tilde{l}_1$, using the $f$ from
Eq. (\ref{charge}).

When the chain collapses ($\tilde{l}_1 \le 1$, the Gaussian value) in a poor solvent away from the critical
point, {\it the chain collects most of its counterions} reducing the effective charge and the
electrostatic penalty, and the two- and three-body interaction terms dominate the free energy ($F_5$).
Here we employ the {\it neutral chain} approximation in which we assume $f \simeq 0$ or
$\alpha=f_m$, and only $F_5$ (with the fourth term in it being zero) survives.
This approximation breaks down (verified numerically) progressively
closer to the critical point where the electrostatic interactions between monomers become
comparable to the excluded volume and steric interactions. However, away from the
critical point $\partial F_5/\partial \tilde{l}_1=0$, in conjunction with
$\tilde{l}_1 \le 1$ in the collapsed state, yields the expansion factor
\begin{eqnarray}
\tilde{l}_1^c=\left(\frac{3}{2}\right)^{2\over3}\frac{2\pi}{3}
\left(\frac{w_3}{|w|}\right)^{2\over3}N^{-{1\over3}}
=2.274\left(\frac{w_3}{|w|}\right)^{2\over3}N^{-{1\over3}}.
\label{l1t-collapse}
\end{eqnarray}
This implies that $R_g^2 \sim R_g^2(\theta)\tilde{l}_1^c \sim N l_0^2 N^{-1/3}$,
or $R_g \sim N^{1/3}$, which is the compact globule limit. Note that $w$ must be
negative to have a collapse ($\tilde{l}_1 \le 1$). 
Now, exactly at collapse the binodal (equal chemical potential) states are,
respectively, the $f=0$ (collapsed) and $f\ne0$ (expanded). Note that for the present situation, chemical
potential is the free energy per monomer.
Equality of chemical potentials of the components in these
two phases having zero osmostic pressure achieved in equilibrium in a dilute solution yields,
using Eq. (\ref{l1t-collapse}) to determine the free energy at collapse,
\begin{eqnarray}
{1\over{Nk_BT}}&&\hskip -0.5cm \left\{(F_1+F_2+F_4)\Big|_{\alpha=\alpha^c}
-(F_1+F_2+F_4)\Big|_{\alpha=f_m}\right\} \nonumber \\
&+&\frac{4}{9}\left(3\over{2\pi}\right)^3
\frac{\left(1-\Theta_t/T\right)^2}{w_3} = 0
\label{collapse-eq1}
\end{eqnarray}
where, $\alpha^c$ satisfies Eq. (\ref{charge}), and the dependency of
$w$ on $T$ is given as
$w=\left(1-\Theta_t/T\right)$. Considering the temperature dependency of the
Bjerrum length, $\tilde{l}_B=C_1/T$, simultaneous solutions of $T$ and $\alpha^c$ in
Eqs. (\ref{charge}) and (\ref{collapse-eq1})
give us the temperature and the
effective charge of the expanded chain at collapse.

\begin{figure}[h]
\includegraphics[width=8.2cm]{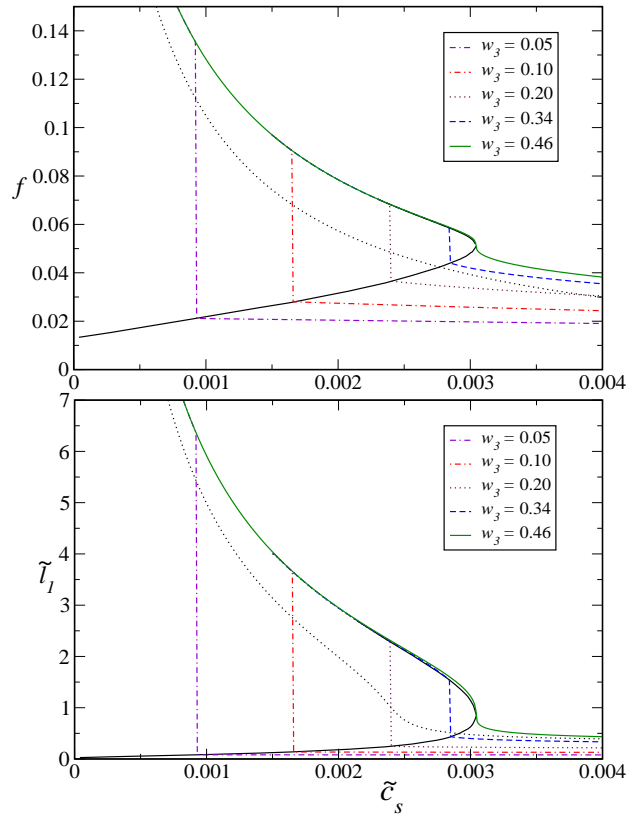}
\caption{\label{fig2}Salt induced collapse of a fully ionizable polyelectrolyte chain.
Fixed parameters are: $N=1000$, $\tilde{\rho}=0.0005$, $\delta=3.0$, Ionizability
$f_m=1.0$, $\Theta_t =400$K, $T=300$K, and therefore, $\tilde{l}_B=3.0$, and $w= -0.333$.
Critical values are: $\tilde{c}_{s1}^\star=0.00305$, $w_3^\star=0.46$,
$\tilde{l}_1^\star=0.9$, and $f^\star=0.051$. Dotted curves are the
analytical result [Eq. (\ref{charge})] in the expanded state.}
\end{figure}

The above approximations which simplify the analysis by decoupling the
chain free energy with rest and allow
us to obtain the closed form expression  of charge [Eq. (\ref{charge})] fail
near the critical point where $f$ is substantially higher than zero. Therefore, to
obtain the critical point and the phase boundaries, we 
minimize the total free energy (sum of all five
components) numerically with respect to both variables (i.e., $f$ and $\tilde{l_1}$).
The critical point and the phase boundaries for the temperature induced collapse of
a fully ionizable chain in salt-free
conditions are presented in Fig.~\ref{fig1}. The unique feature of this
calculation is to take temperature as the only independent variable, thus rendering
$w$ and $l_B$ as dependent functions through, respectively, the relations
$w=1-\Theta_t/T$, and $\tilde{l}_B=18000/\epsilon T l_0$, assuming that the Bjerrum length
in water at room temperature is $\sim 0.75$ nm, and the Kuhn length ($l_0$) is
$\sim 0.25$ nm [keeping in mind a polymer of type sodium polystyrene sulfonate (NaPSS)].
The figures are representations of phase transitions with the theta-temperature and dielectric constant
as reference points, and the temperature a general experimental variable.
Notable is the qualitative similarity of the charge ($f$) and size ($\tilde{l_1}$) phase
boundaries. There can not be a coexistence of expanded and collapsed states
for the same effective charge of the chain. In other words, {\it at equilibrium the chain can
have only one size for one value of its effective charge}. Further,
the critical size of a fully ionizable chain at salt free conditions is larger
than its Gaussian value ($\tilde{l_1}=1$) due to intra-chain electrostatic repulsion, whereas for 
a neutral chain it is smaller ($\tilde{l_1}^\star=0.45 < 1$).  
Gradual reduction of the ionizability ($f_m$) leads to an
increasing critical temperature and a decreasing critical size, for which the
neutral chain values\cite{mut-84} are recovered for $f_m=0$. The effect of ion-pairs\cite{mut-04}
that reduces the two-body interaction parameter only slightly without
significantly affecting the phase boundaries is ignored. The phase boundary clearly suggests that counterion
condensation and chain collapse are mutually cooperative processes in the case of a varying
Coulomb strength. Lowering temperature in the
expanded state helps counterion adsorption which reduces the effective charge and hence
the resistance to collapse. Once the collapse occurs, charge becomes negligible with rampant counterion adsorption
minimizing the electrostatic energy penalty.
One may contrast this cooperative behavior with a purely collapse induced
collection of counterions\cite{lohetal-08}, where only the solvent quality (for the mixture of 1-propanol 
and 2-pentanone containing quarternized poly-2-vinylpyridine) is reduced keeping
the the Coulomb strength (Bjerrum length times dielectric constant) virtually unchanged. 
Similar measurements of size and ionic conductivity as the temperature is varied would 
validate the major concepts presented in this letter.

In Fig.~\ref{fig2} we present the phase boundaries for the salt
induced collapse of a fully ionizable PE at $300K$, at which
the solvent is poor (note, $\Theta_t=400K$). This shows that room temperature collapse of polymer of 
type NaPSS in water is
possible in the presence of salt but unlikely at salt-free conditions (Fig.~\ref{fig1}) unless solvent poorness is
drastically enhanced\cite{lohetal-08}. 
There are remarkable qualitative similarities between salt and
temperature induced collapse (Fig.~\ref{fig1}).  
The main ideas of the phase transitions, arising from the coupling between 
counterions and polymer conformations, are also valid for the volume transitions of PE gels and 
brushes. New experiments monitoring the counterion density inside the gels and brushes will stimulate a 
fundamental understanding of this important class of soft materials. One finally notes that computer 
simulations\cite{dobetal-lyuetal-limall-kirall-96-99} ignore the dielectric heterogeneity present in real 
experimental polymeric systems. Equivalently, a low value of $\delta$ in our theory for which 
the counterion adsorption is weak raises the possibility of a necklace-like structure. Experiments\cite{lohetal-08}, 
however, see rough spherical objects with substantial incorporation of counterions within them, as predicted by our model.

We thank Rajeev Kumar for stimulating discussions. This work was supported
by NIH Grant No. 5R01HG002776 and
National Science Foundation (NSF) Grant No. 0605833. 

\noindent $\star$ {\small corresponding author:
muthu@polysci.umass.edu




\end{document}